\pdfoutput=1
\documentclass[a4paper]{tufte-handout}
\usepackage{amsmath}
\usepackage{graphicx}
\setkeys{Gin}{width=\linewidth,totalheight=\textheight,keepaspectratio}
\graphicspath{{Figures/}}
\usepackage{booktabs}
\usepackage{units}
\usepackage{fancyvrb}
\fvset{fontsize=\normalsize}
\usepackage{multicol}

\usepackage{amsfonts}
\usepackage{natbib}
\def\adx#1:#2\par{\par\halign{\hskip #1##\hfill\cr #2}\par}
\def\first{$1^{\mathrm{st}}$}

\def\teff{T_{\rm eff}}
\def\mast{M_\ast}
\def\rsol{R_\odot}
\def\msol{M_\odot}
\def\lsol{L_\odot}

\def\nabad{\nabla_{\mathrm{ad}}}
\def\nabrad{\nabla_{\mathrm{rad}}}

\def\H1{^1\mathrm{H}}
\def\He3{^3\mathrm{He}}
\def\He4{^4\mathrm{He}}
\def\C12{^{12}\mathrm{C}}
\def\N14{^{14}\mathrm{N}}
\def\O16{^{16}\mathrm{O}}

\def\unity{ \hbox{1\kern-.23em l} }
\def\zero{ \hbox{0\kern-.23em |} }
\def\field{ \hbox{I\kern-.23em K} }

\def\braket #1.#2.{\langle #1 \vert #2 \rangle}

\def\O{\mathcal{O}}
\usepackage[fulloldstylenums]{kpfonts} 
\title{The Thermal Pulses of Very-Low-Mass Stars}
\author[]{Alfred Gautschy, CBmA Liestal and ETH-Bibliothek Z\"urich}

\begin{document}
\maketitle

\begin{abstract}
  \noindent Very-low-mass stars can develop secularly unstable
  hydrogen-burning shells late in their life. Since the thermal pulses
  that go along are driven at the bottoms of very shallow envelopes,
  the stars' luminosities and effective temperatures react strongly
  during a pulse cycle.  Towards the end of the Galaxy's stelliferous
  era, the hydrogen-shell flashing very-low-mass single stars should
  inflict an intricate light-show performed by the large population of
  previously inconspicuous dim stars. Unfortunately, this natural
  spectacle will discharge too late for mankind to indulge in. Not all
  is hopeless, though: In the case of close binary-star evolution,
  hydrogen-shell flashes of mass-stripped, very-low mass binary
  components can develop in a fraction of a Hubble time. Therefore,
  the Galaxy should be able put forth a few candidates that are going
  to evolve through a H-shell flash in a humanity-compatible time
  frame.
\end{abstract}

\bigskip

\section{The Evolution of Single Very-Low-Mass Stars}

\begin{marginfigure}[4cm]
\includegraphics{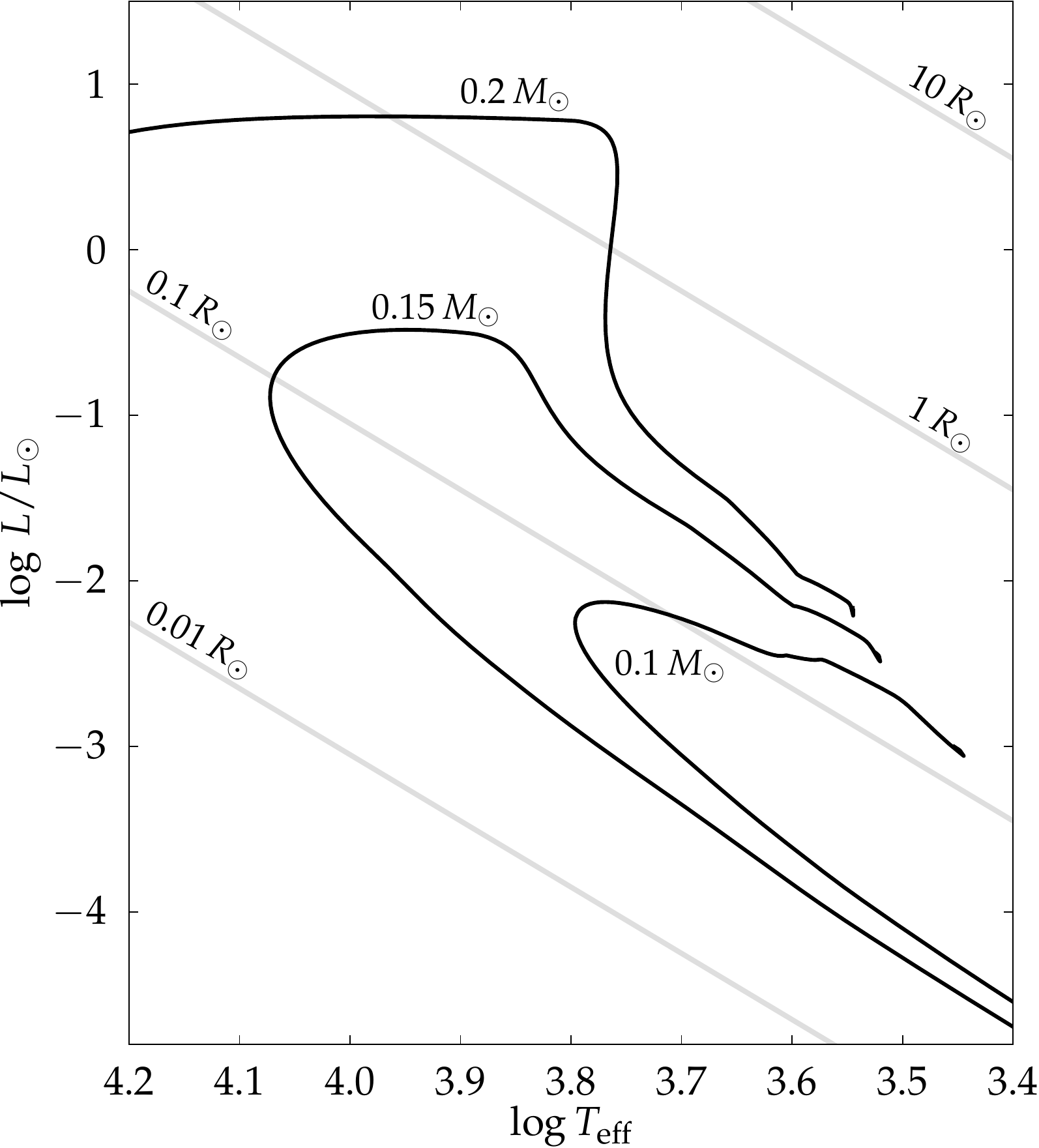}
\caption{Examples of VLM evolutionary tracks~--~$0.1, 0.15$, and $0.2
         \msol$ with $Z=0.02$~--~on the HR plane.
        }
\label{fig:vlmhrd}
\end{marginfigure} 

\newthought{Red dwarfs} in the mass range of about $0.08 - 0.45\,
\msol$, referred to as very-low-mass (VLM) stars hereafter, are
massive enough to start nuclear burning of hydrogen but \emph{not}
massive enough to enter the stage of helium burning; these stars do
not rank highly as daring astrophysical objects. Indeed, most or all
of their life is quiet and dim, or put otherwise: unspectacular. For
most of the nuclear lifetime of the VLM stars, which exceeds the
current age of the Universe by a large margin, they stay put close to
the main sequence on the low luminosity and low effective-temperature
corner of the HR diagram. In particular the low-mass fraction of the
red dwarfs, which remains essentially fully convective during most of
the main-sequence phase, feeds the nuclear-burning core with fresh
fuel from throughout the whole stellar volume for a long time, ranging
from about 100~Gyrs for $\approx 0.4 \msol$ to more than 1000~Gyrs for
$\approx 0.15 \msol$ stars.
\marginnote{Numerical data cited in the text without explicit
  reference refer either to results from own computations with the
  MESA stellar evolution code (cf. Appendix) or are considered to be
  part of the current astronomical canon.}

Only those VLM stars that develop radiative cores at some stage during
central hydrogen burning ascend the \first~giant branch noticeably,
i.e. VLM stars with $M_\ast \gtrsim\,0.15 \msol$ inflate at least
somewhat.  Since stars with $M_\ast \lesssim 0.45 \msol$ do not ignite
core helium burning, they leave the giant branch once the envelope
mass (i.e. the mass above the slowly outward-eating hydrogen burning
shell) drops below a critical value. Both, the reduction of envelope
mass and the ever increasing stellar radius of the stars ascending the
red-giant branch both contribute to the reduction of the total
pressure at the hydrogen-burning shell so that eventually $\teff$ must
increase in order to be able to transport to the surface the energy
generated at the H-shell.  The luminosity at which the VLM stars divert from
the fiducial giant branch correlates with the total stellar mass:
The higher the total mass, the brighter the giant upon departure from
the giant branch; at the same time, the remaining relative envelope
mass diminishes.

After leaving the giant branch, the VLM stars cross the HR diagram at
about constant luminosity and evolve to effective temperatures
exceeding $10^4$~K, see e.g. Fig.~\ref{fig:vlmhrd}. Again, the maximum
temperature correlates with the stellar mass in such a way that higher
mass stars achieve higher maximum $\teff$. The crossing at constant
luminosity ($\Delta t = {\cal O}(10^8 - 10^9)$~years for $0.2\,\msol$
and ${\cal O}(10^7)$~years for $0.35\,\msol$, subject to composition
and the particular microphysics invoked) and the early cooling along
the characteristic cooling helium white-dwarf branch are quick
compared with the earlier stages of evolution. The horizontal
evolution at constant luminosity terminates once the previously
generated luminosity, in accordance with the core-mass~--~luminosity
relation, can be maintained in the ever less massive hydrogen-burning
shell which already gets cooler as the star heats up at the
surface.

With the luminosity support of the weaker-getting hydrogen-burning
shell the VLM stars approach the helium-white-dwarf cooling region.
Not all VLM stars evolve monotonously through the early degenerate
cooling phase; for some, the hydrogen shells undergo one or several
thermal flashes. Since the secularly unstable hydrogen-burning shells
lie at the bottoms of only very shallow envelopes the respective stars
react sensitively and embark on complicated trajectories on the HR
diagram.

\section{Hydrogen Shell Flashes in VLM Stars}

The central part of this exposition focuses on the \emph{thermally
  pulsing} VLM stars; we are going to look into the physical
circumstances of the H-shell flashes and their r{\^o}le in stellar
astronomy of single VLM stars and as components in close binary
systems as observable currently.

Figures~\ref{fig:z02tracks} to \ref{fig:z02_diff_tracks} all show
evolutionary tracks of selected model sequences, first with changing
heavy element abundance, and then illustrating the effect of elemental
diffusion. Each figure contains evolutionary tracks of stars just above
and below the secularly unstable mass range in gray; two selected
thermally pulsing representants in between illustrate topology and
extent of the tracks that result as a reaction on the H-shell
flashes.  

\subsection{Non-diffusive star models}
To start, we focus first on the results of \emph{simple} evolutionary
computations, meaning in the present context neglecting elemental
diffusion. Independent of the stars' metallicity, there exists always
a range of stellar masses within which the physical circumstances were
favorable for the stars to develop secularly unstable H-shell burning
early on the white-dwarf 
cooling track.\sidenote[][-3cm]{Historically, \citet{Kippenhahn1968}
  were the first to encounter a hydrogen-shell flash in a low-mass
  star. The authors were interested in following the long-time
  evolution of an interacting binary system, simulated with a
  single-star evolution code.  To learn about later work on this topic
  see e.g. \citet{Driebe1998}, \citet{Sarna2000},
  \citet{Benvenuto2005} and references therein.  }
As a consequence, the secularly unstable models underwent one or a few
hydrogen flashes, which forced them to loop extensively on the HR
plane. The \emph{magnitude} and \emph{topology} of the excursions away
from the evolutionary tracks as traced out by stars burning hydrogen
quietly depended on stellar mass as well as on chemical
composition. The particularities of the microphysics treatment, as
well as the chemical composition  influenced the mass range of
stars that underwent hydrogen shell flashes.

\begin{figure}
	\label{fig:z02tracks}
	\includegraphics{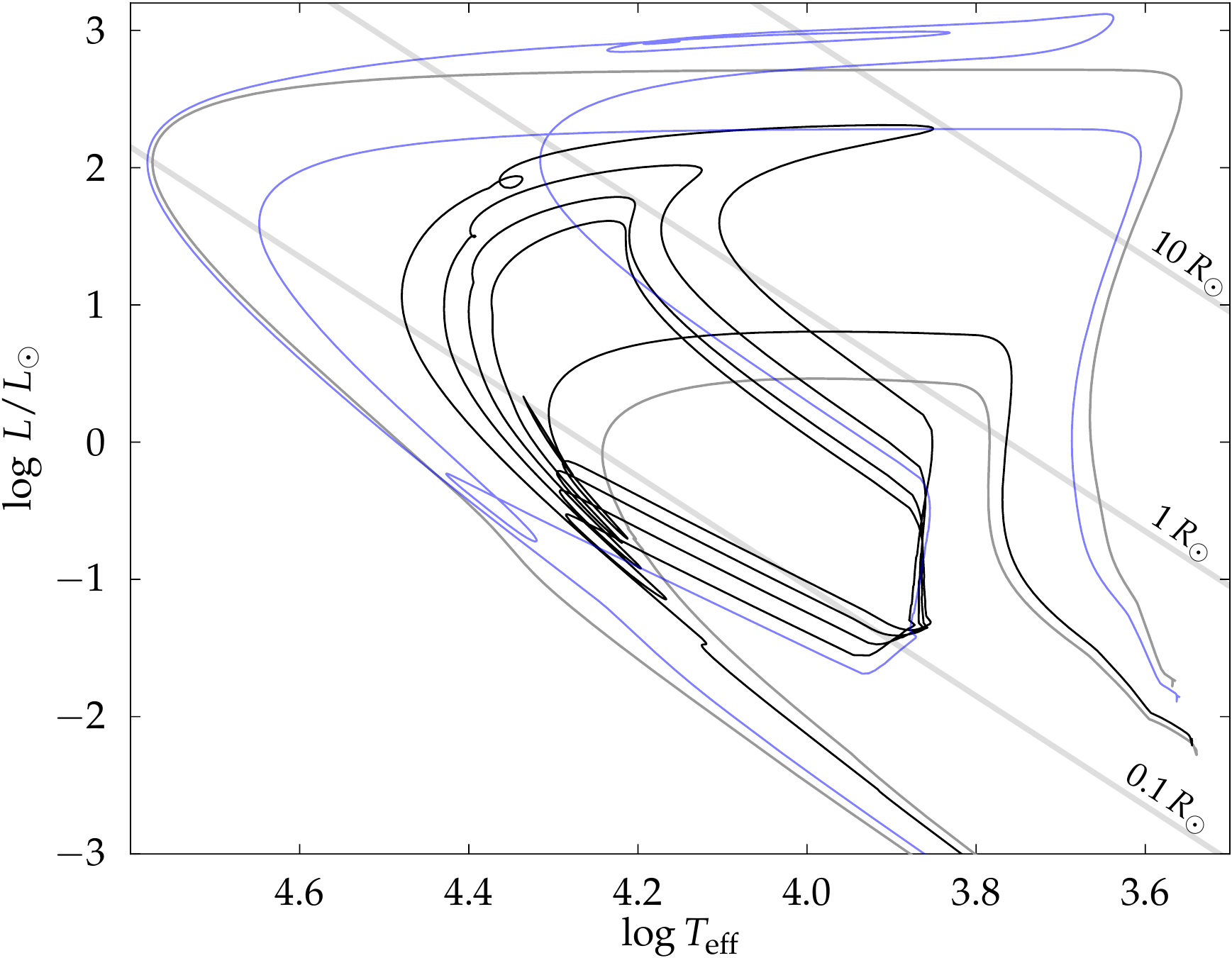}
	\caption{Evolutionary tracks on the HR plane of $Z=0.02$ VLM
          models. The two gray tracks are traced by a $0.185\,\msol$
          model on the low-mass and a $0.35\,\msol$ model on the
          higher-mass side; these two stellar masses bracket the
          domain of the thermal pulsing instability. For illustration,
          the fine black line, including four thermal pulses, shows
          the track of a $0.2\,\msol$ model. The blue line is traced
          out by a $0.3\,\msol$ model, which lives through only one
          thermal pulse of the H-burning shell anymore.  }
\end{figure}

In the case of solar-like $X=0.7, Z=0.02$ models, secularly unstable
H-shells were encountered in the range from $0.19$ to $0.32 \msol$.
The thermal flashes are most vigorous close to the lower mass
boundary. Between $0.19$ and $0.2 \msol$ the model sequences passed
through four hydrogen-shell flash cycles. From $0.26$ to $0.32 \msol$,
only one single hydrogen flash took place. Selected evolutionary
tracks illustrate the situation in Fig.~\ref{fig:z02tracks}.

Of the four tracks included in Fig.~\ref{fig:z02tracks}, the two gray
ones delineate roughly the upper ($0.35\,\msol$) and lower
($0.185\,\msol$) mass boundary in between of which the unstable
hydrogen shells were encountered. The black locus is indicative of a
multi-flash cycling VLM star~--~here a $0.2\,\msol$ model sequence
evolving through four cycles. The blue-colored evolutionary locus shows
the example of a star close to the upper mass boundary for unstable
hydrogen burning; these stars, although only living through a single
pulse can pass through a very puffed-up born-again red-giant phase
before they settle again on the quiet white dwarf cooling track. For
better illustration of the amount of radius growth during hydrogen
flashes, lines of constant radii are added to the plot. Notice that
during the flash-induced excursions on the HR plane, the radii of
these VLM stars can grow temporarily ten- to hundredfold.  

\begin{figure}
	\label{fig:z001tracks}
	\includegraphics{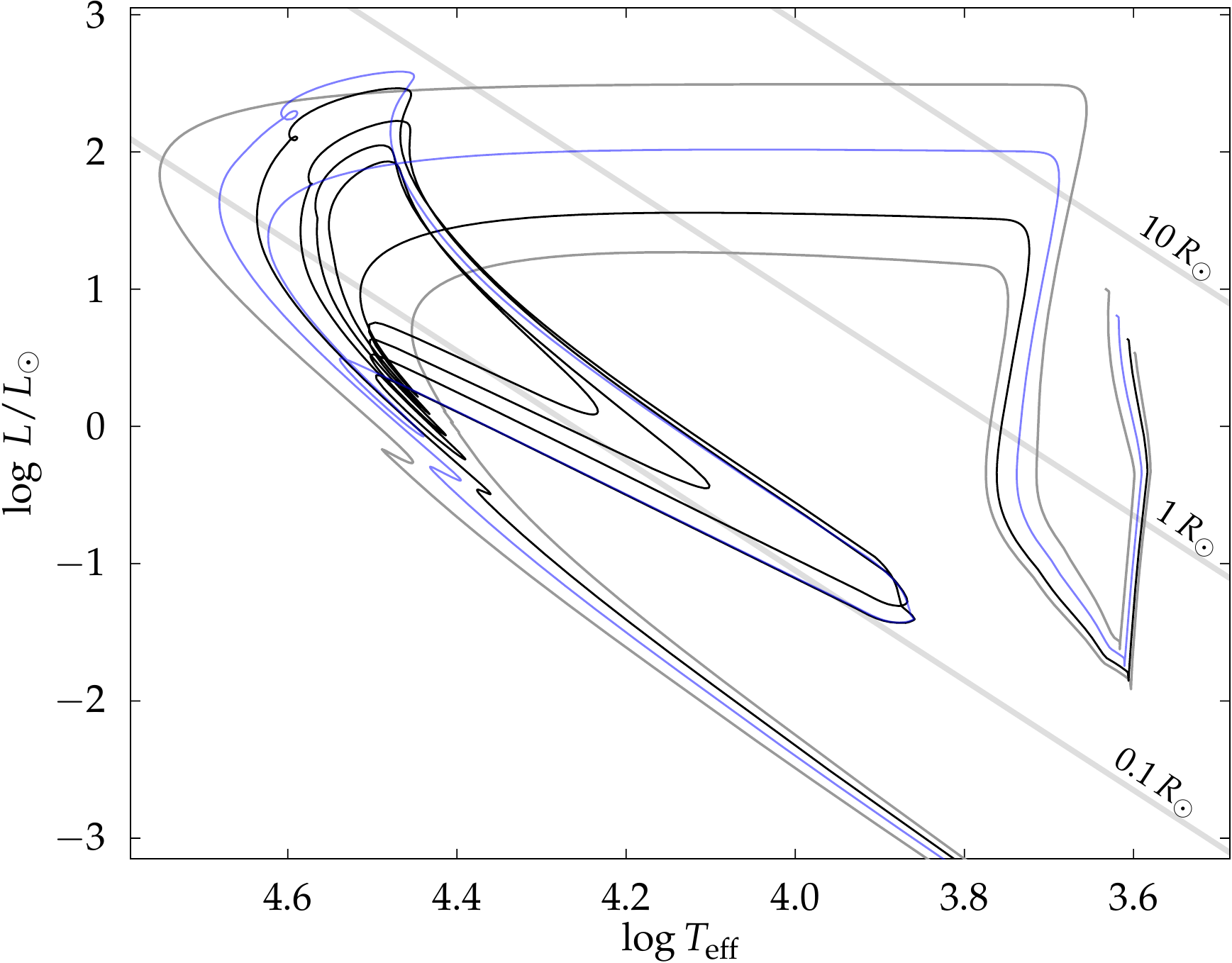}
	\caption{Evolutionary tracks of $Z=0.001$ VLM models on the HR
          plane. The two gray tracks are traced out by a $0.24\,\msol$
          model on the low-mass and a $0.355\,\msol$ model on the
          higher-mass side. These two stellar masses bracket the
          domain of the thermal pulsing instability. For illustration,
          the fine black line, including four thermal pulses, shows
          the track of a $0.26\,\msol$ model. The blue locus is traced
          out by a $0.3\,\msol$ model, which lives through only one
          thermal pulse of the H-burning shell anymore.  }
\end{figure}

Figure~\ref{fig:z001tracks} is closely related to
Fig.~\ref{fig:z02tracks} but shows evolutionary tracks on the HR plane
of selected star models with $X=0.757, Z=0.001$. The tracks are shown from
their starting as thermally contracting, homogeneous gas spheres
at the Hayashi line which then approach the main sequence before
they set out on the same evolutionary voyage as the models in
Fig.~\ref{fig:z02tracks}.

The metal-poor VLM models showed also a range of stellar masses with
secularly unstable hydrogen shells. Starting at $0.25 \msol$, up to
$0.26 \msol$, with four thermal flash cycles, the temporarily unstable
hydrogen-burning shells persisted up to $0.35
\msol$. Figure~\ref{fig:z001tracks} shows in gray quietly H-burning
model sequences with $0.24$ and $0.355 \msol$, jammed in between is a
model evolving through four hydrogen-shell flashes (in black) and a
model sequence (in blue) close to the upper mass boundary encountering
a single thermal flash.

Phenomenologically, the major difference between the $Z=0.001$ and the
$Z=0.02$ models is the reduced radius change encountered during a
hydrogen shell flash by the metal-poor star models. Reminiscent of
other situations in stellar astronomy, metal poor star tend to prefer
the bluer parts of the HR diagram as compared to the metal richer
populations. Most pronounced we see this manifested in the lacking red
noses~--~compare Fig.~\ref{fig:z001tracks} with
Fig.~\ref{fig:z02tracks}~--~of the evolutionary tracks at high
luminosity during the hydrogen flashes. The metal poor VLM stars do
not produce born-again red giants but only bright blueish giants.

\subsection{Models with elemental diffusion}
The nuclear-evolution timescale of single VLM stars is very long,
exceeding the current age of the Universe. Hence, these stars are
candidates for diffusion to play a r{\^o}le for structure and
evolution as soon as radiative regions develop in their interiors. For
the illustrative purposes of this exposition, we computed a set of
$Z=0.02$ models to get a rough impression of the effect of
elemental diffusion on existence and prevalence of the H-shell flashes
and surface character of diffusive model stars.

\begin{figure}
	\label{fig:z02_diff_tracks}
	\includegraphics{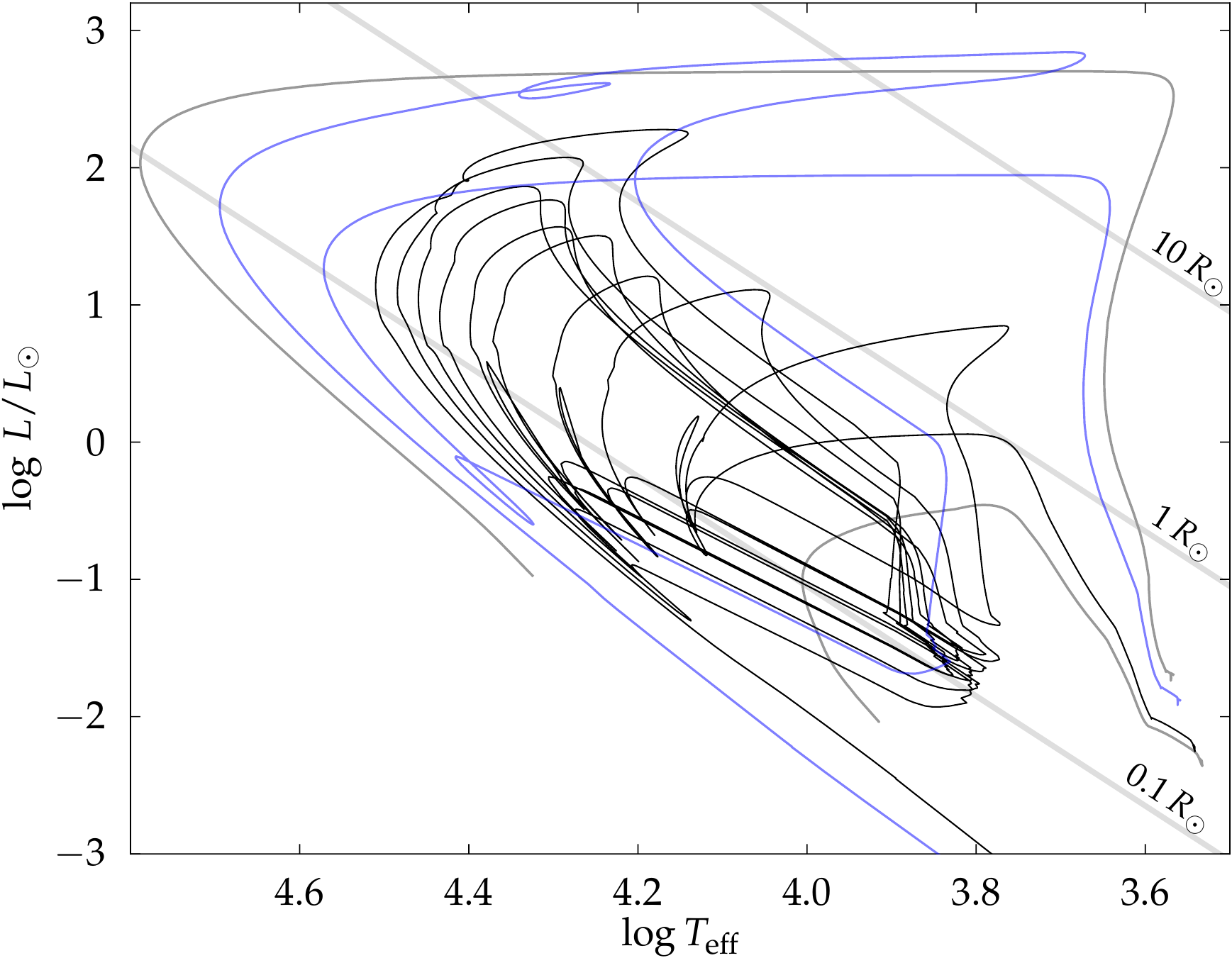}
	\caption{Evolutionary tracks of $Z=0.02$ \emph{diffusive} VLM
          models on the HR plane. The two gray tracks are traced by a
          $0.17\,\msol$ model on the low-mass and a $0.37\,\msol$
          model on the high-mass side of the secular instability
          domain. For illustration, the thin black line, including
          nine thermal pulses, shows the track of a $0.19\,\msol$
          model. The blue line is traced out by a $0.29\,\msol$ model,
          which lives through only one thermal pulse of the H-burning
          shell anymore.  }
\end{figure}
Figure~\ref{fig:z02_diff_tracks} shows, in the same spirit as
Figs.~\ref{fig:z02tracks} and \ref{fig:z001tracks}, selected
evolutionary tracks of star models with $Z=0.02$, which were computed
including elemental diffusion. The gray lines belonging to the $0.17
\msol$ and the $0.37 \msol$ tracks trace out the lower and upper
boundary in between of which stars undergo hydrogen-shell flashes. In
contrast to the diffusion-free models, thermal flashes are more
numerous when elemental diffusion affects the stars' abundance
stratification. The maximum number of 9 thermal pulses was encountered
in the case of the $0.19 \msol$ model star, its track is shown as the
black line in Fig.~\ref{fig:z02_diff_tracks}. The example of a star
with $0.29 \msol$, close to the upper mass-boundary of H-shell
instability, undergoing only a single hydrogen shell flash with an
extensive excursion into the red-giant region is also included in
Fig.~\ref{fig:z02_diff_tracks}; its track is shown with the blue line.
Comparing the blue tracks of Fig.~\ref{fig:z02tracks} and
Fig.~\ref{fig:z02_diff_tracks} makes clear that evolutionary tracks of
the single H-shell flash stars do not diverge strongly; hence, the
effect of elemental diffusion on the phenomenology of the tracks is
minimal.

\begin{figure}
	\label{fig:z02_diff_HTPs}
	\includegraphics{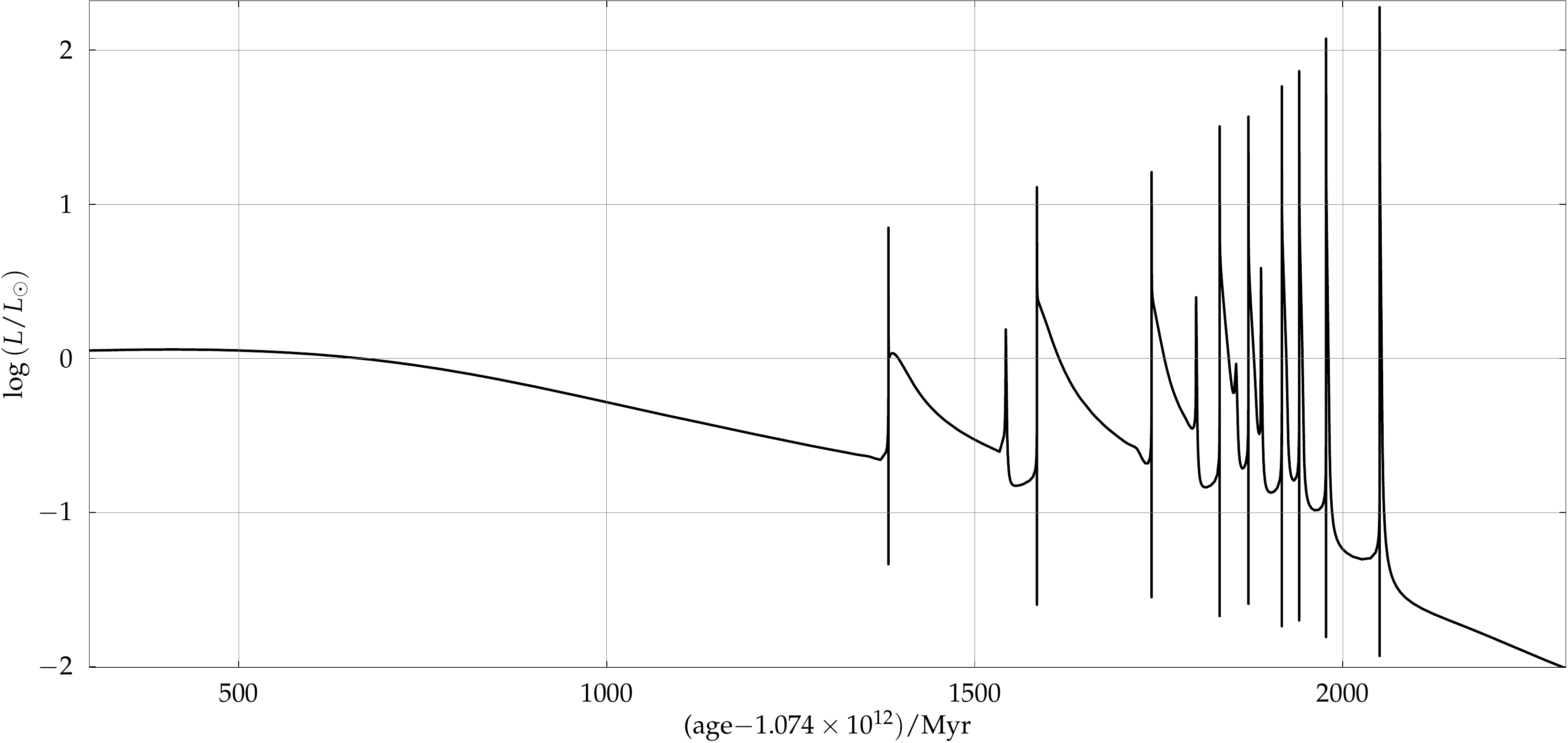}
	\caption{Temporal evolution of the luminosity of a
          $0.19\,\msol,\, Z=0.02$ model star around the evolutionary
          stage of secularly unstable hydrogen shell burning. For
          convenience, the plotted age of the star is offset by
          $1.074\times10^{12}$; on this scale, giant-branch evolution
          of the star terminates around $500$. }
\end{figure}
The bolometric lightcurve of the thermally pulsing $0.19\,\msol$ star
of Fig.~\ref{fig:z02_diff_tracks} is displayed in
Fig.~\ref{fig:z02_diff_HTPs}.  The abscissa measures the time in
mega-years relative to the arbitrary, but convenient epoch at
$t=1.074\times 10^{12}$~years, measuring roughly the age of the model
star on the top of the \first~giant branch. The lightcurve shown in
Fig.~\ref{fig:z02_diff_tracks} makes it clear that the thermal pulses
induced by the hydrogen-burning shell are quite irregular, much more
so than those of He-shell flashes. The interpulse-period can range
from 202~Myrs (between pulse 1 and 2) to 24~Myrs (from pulse 6 to 7).
The amplitude of the pulses, though, grow continuously. During the
first pulse, the luminosity rises by about $100\,\lsol$; then, during
the last pulse, the luminosity amplitude grows to about
$10\,000\,\lsol$.
\marginnote[-1cm]{The He-shell flashes that occur in stars along the
  asymptotic giant branch are computed to be much more regular
  \citep[e.g.][Chapter 18.3]{Iben2013_2}; they even follow a reliable
  correlation of core mass and inter-pulse period
  \citep{Paczynski1975}.}
The lightcurves of the thermal-pulse phase reflects also the
complicated loci traced out by the model star on the HR plane (see
Fig.~\ref{fig:z02_diff_tracks}). We observe considerable substructure
in a thermal pulse; an aspect which got some attention in \citet{Driebe1999}.
\sidenote{The stellar modeling used in \citet{Driebe1999} did not
  include elemental diffusion; the range of stellar masses found to
  encounter H-shell flashes, $0.21 - 0.3\,\msol$, compares favorably
  with our diffusion-less model sequences.}
The substructures within a particular pulse cycle do, however, not
necessarily repeat in subsequent cycles. It can even be tricky, based
on the luminosity variation alone, to determine when a cycle has
finished. Analyzing the track on the HR plane, though, resolves this
problem.

\begin{figure*}
\label{fig:z02_Kip_NucBurn}
\centering
\includegraphics[width=0.46\linewidth]{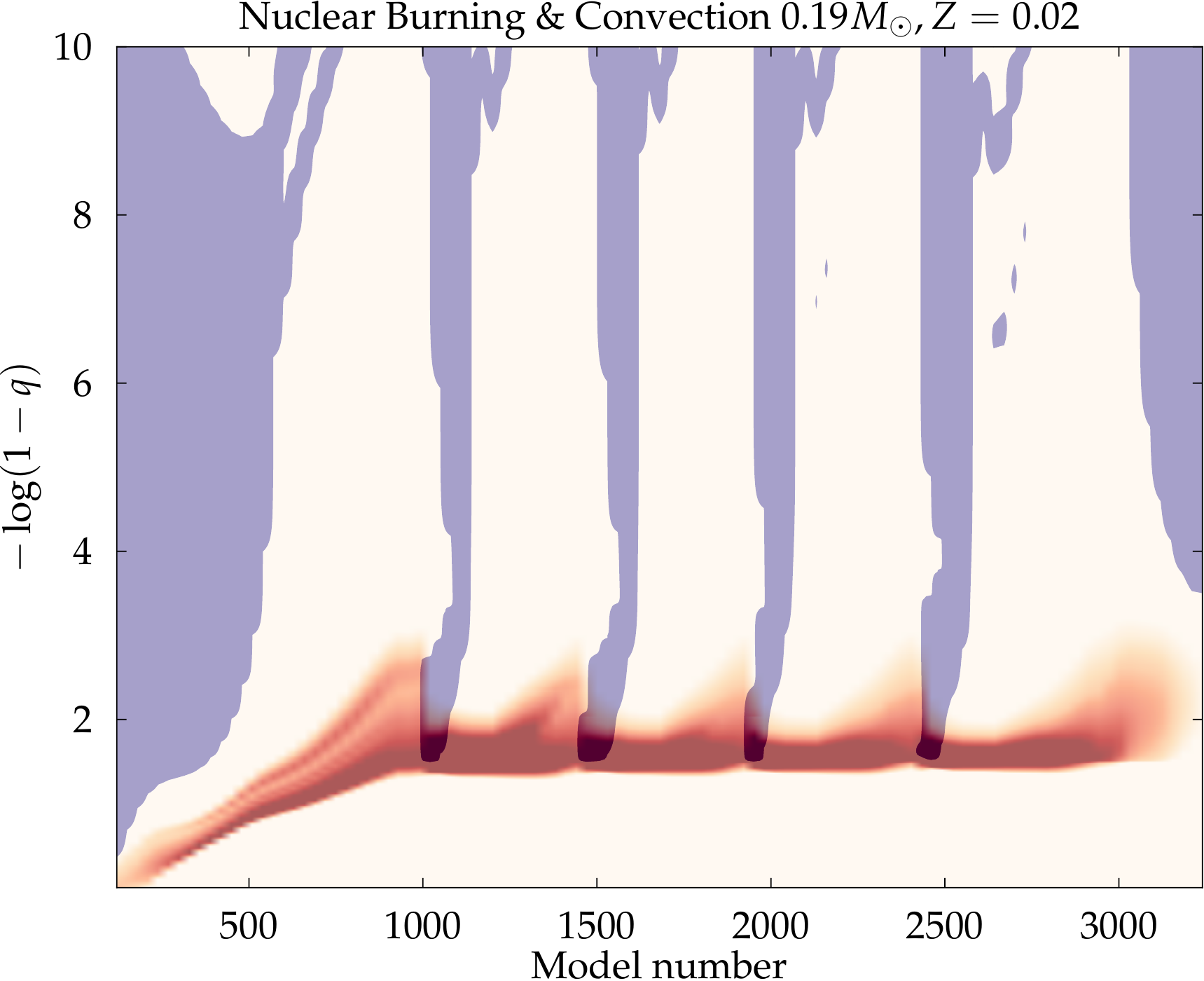}
\hspace{0.3cm}
\includegraphics[width=0.46\linewidth]{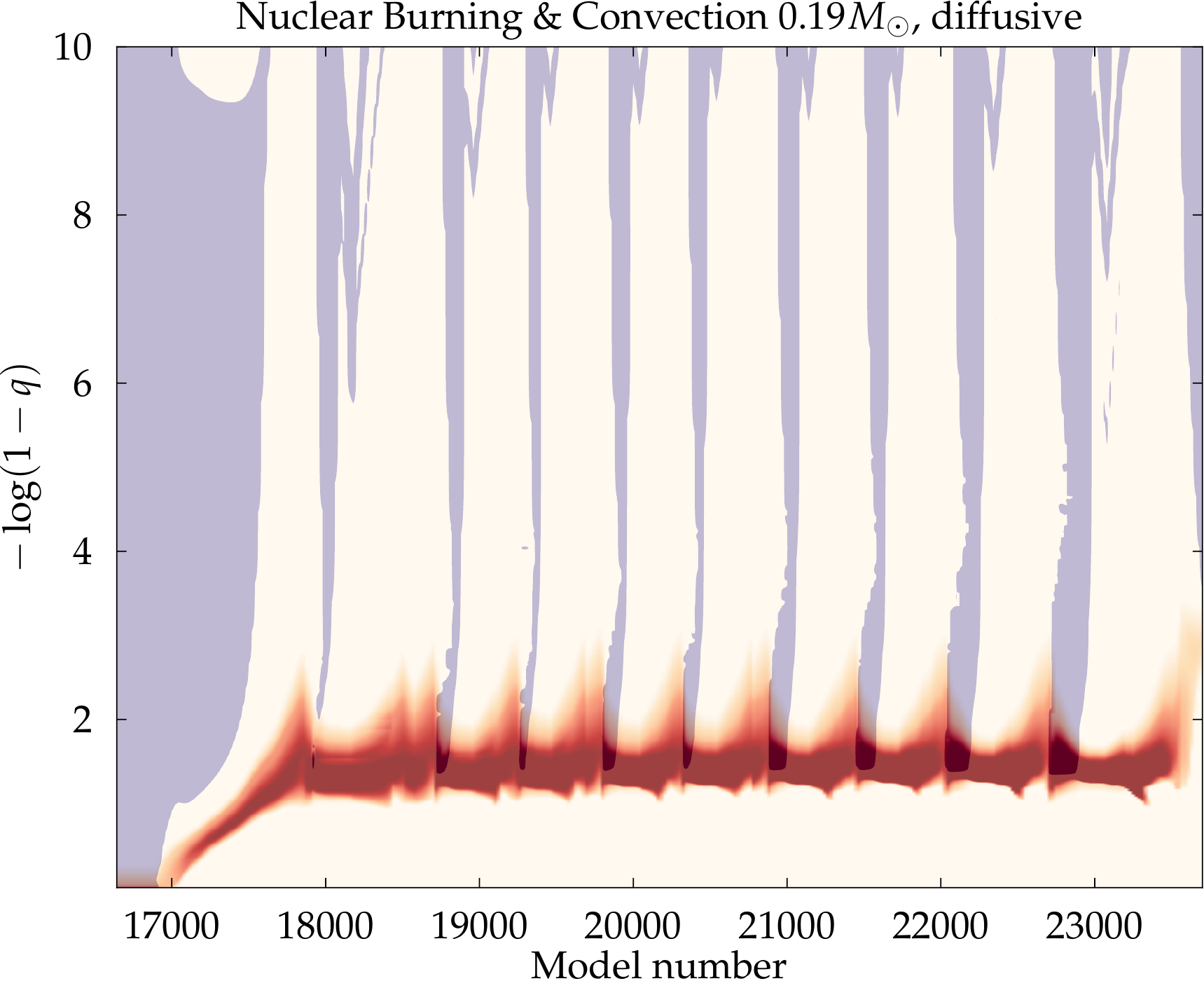}
\caption{Kippenhahn diagrams tracing out the convection zones (blue)
  and the strength of nuclear burning (red) of $0.19\,\msol$ model
  stars. The left panel shows the case of a \emph{non-diffusive} model, the
  right panel contains the result from modeling the star inclusive
  \emph{elemental diffusion}.
}
\end{figure*}

\begin{figure*}
\label{fig:z02_Kip_Xabun}
\centering
\includegraphics[width=0.47\linewidth]{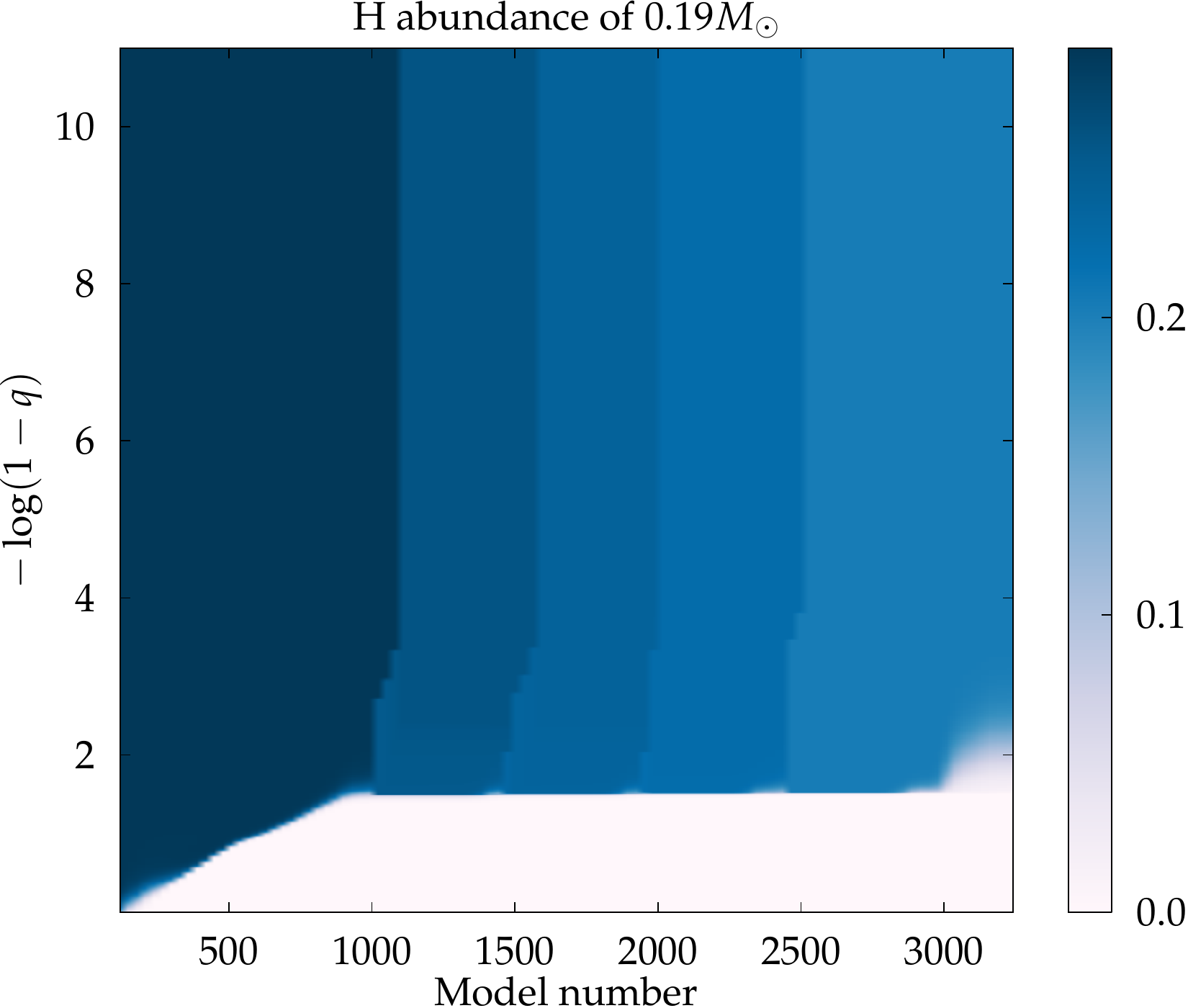}
\hspace{0.3cm}
\includegraphics[width=0.47\linewidth]{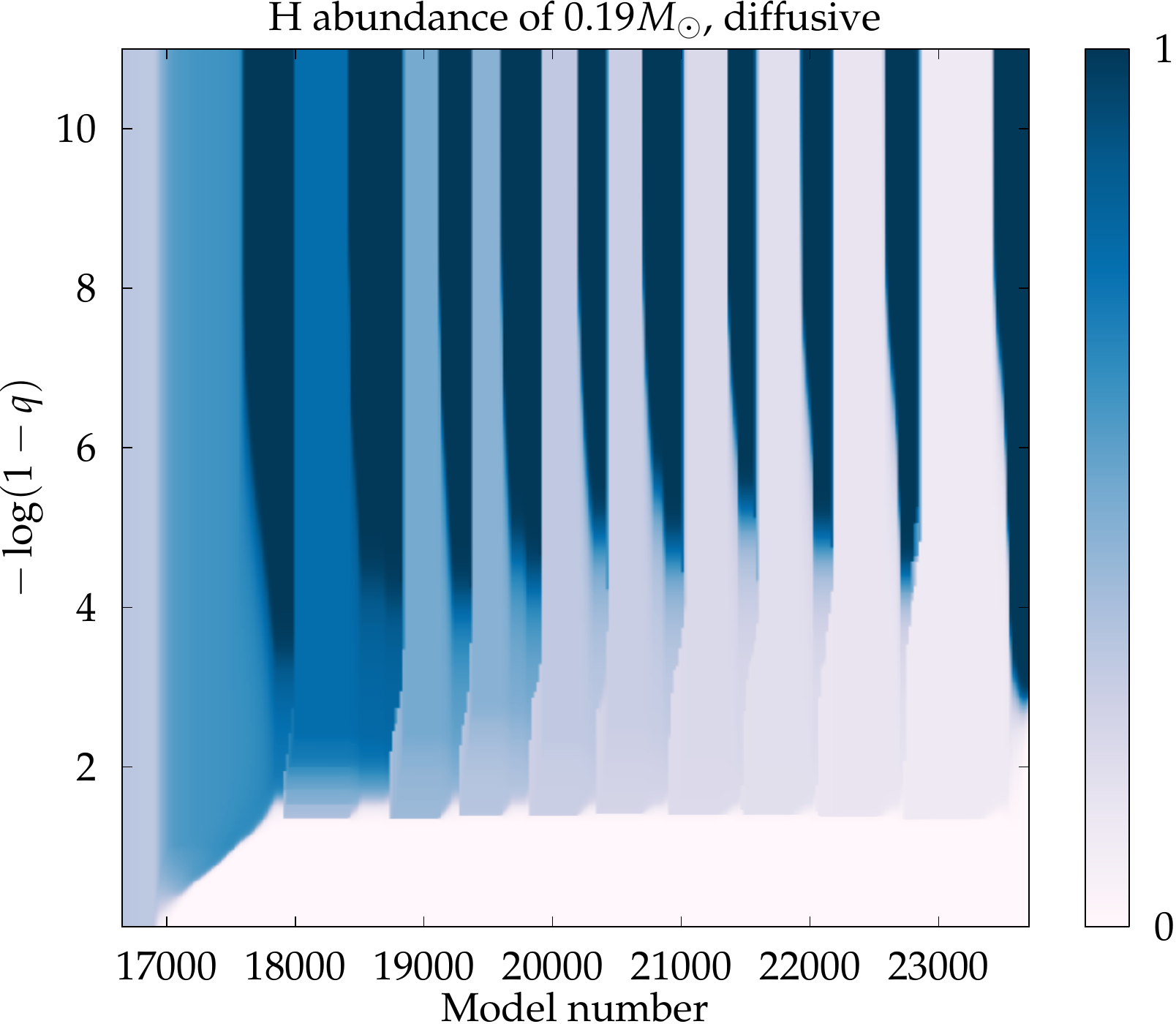}
\caption{Kippenhahn diagrams showing the spatial evolution of 
  the hydrogen abundance of $0.19\,\msol$ model stars.}
\end{figure*}

\newthought{The Kippenhahn diagrams} of
Fig.~\ref{fig:z02_Kip_NucBurn}, plotting $-\log(1-q)$ with
$q=m/\mast$, i.e. measuring the remaining relative mass lying above a
given mass coordinate, versus model number to appropriately but
non-monotonously stretch the time axis, show the behavior of a
non-diffusive $0.19\,\msol$ model in the left and a diffusive model of
the same mass in the right panel.  The chosen model numbers in the
panels essentially zoom in to the evolutionary window between the end
of core hydrogen burning and the termination of hydrogen shell flashes
when also nuclear energy generation dies out.  Model numbers of the
non-diffusive model stars shown in the left panels of
Figs.~\ref{fig:z02_Kip_NucBurn} and \ref{fig:z02_Kip_Xabun} can be
related to physical ages using the table on the margin. For the
diffusive case, correlating the thermal flashes in
Fig.~\ref{fig:z02_diff_HTPs} with the associated convection fingers in
Fig.~\ref{fig:z02_Kip_NucBurn} does the job.
\begin{margintable}
\begin{tabular}{r r}
\toprule
Model&  age/$10^{12}$~yrs \\
\midrule
 500 & 1.15053 \\
1000 & 1.15082 \\
1500 & 1.15084 \\
2000 & 1.15086 \\
2500 & 1.15089 \\
3000 & 1.15131 \\
\bottomrule
\end{tabular} 
\end{margintable} 
The shades of red in Fig.~\ref{fig:z02_Kip_NucBurn} indicate nuclear
burning with deeper red meaning higher specific nuclear energy
generation rate; the particular magnitudes are of no interest
here. Blue regions map out convection zones (according to
Schwarzschild's criterion); in the remaining pale the background,
energy is transported by radiation diffusion. The big blue patches on
the left of both panels indicate the fully convective red dwarfs and
the red-giant phase with its very deep convective envelopes on top of
the hydrogen-burning shell.  The thin convective branches, in the
region $8\lesssim -\log(1-q)\lesssim 10$, growing out of the
dominating vertical convection trunks are the robust convection zones
induced by partial hydrogen and helium ionization; the superficial
convection zones persist even when essentially the whole envelope
becomes radiative when the model stars heat up. Depending on the
prevailing density structure in the stellar envelope, the zones merge
or appear as separate thin convection zones. The patchy convective
regions, appearing only during the late thermal-pulse phase, deeper in
the stars' envelopes ($6\lesssim -\log(1-q)\lesssim 8$) are
attributable to the Fe-bump of the opacity. At sufficiently high
envelope density, the local opacity peak can push up $\nabrad$ to make
the region superadiabatic and hence convectively unstable.

The hydrogen-shell flashes are not only discernible as temporarily
extended and stronger nuclear-burning regions in
Fig.~\ref{fig:z02_Kip_NucBurn} but also by the ensuing deep convection
zones which reach into the nuclear burning region of the star models
(except for the first pulse in the diffusive model). Such deeply
penetrating convection means that nuclearly processed material
(CNO-cycle processed in the current case) will be dredged up to the
stellar surface.

For the same models and the same evolutionary window as shown in
Fig.~\ref{fig:z02_Kip_NucBurn}, the temporal evolution of the hydrogen
abundance stratification of non-diffusive (left panel) and diffusive
models (right panel) are illustrated in
Fig.~\ref{fig:z02_Kip_Xabun}. \marginnote{ Notice that the deepest
  blue coloring amounts to different maximum hydrogen abundances in
  the left and the right panel, respectively.} The white bottom
regions of the two panels measure the mass of the degenerate, nuclear
inactive helium core.

In accordance with the convection zones and the nuclear burning, in
particular during the hydrogen-shell flashes, the left panel of
Fig.~\ref{fig:z02_Kip_Xabun} shows a discontinuous hydrogen depletion
in the stars' envelope as the star model passes through its four
thermal pulses. The process starts at the low, spatially homogeneous
value of $X\approx 0.3$ at the end of core hydrogen burning; $X$ is so
low because the $0.19\,\msol$ model stayed fully convective for a
significant fraction of its main-sequence lifetime so that about half
of the star's \emph{total} hydrogen content was consumed during its
main-sequence phase. The envelope hydrogen abundance remains constant
up to the first thermal flash when a fraction is burned up and the
remaining $X$ abundance is homogenized via convection through the
stellar envelope. With each further thermal flash the star's hydrogen
abundance gets reduced and homogenized; this process produces the
patchy sequence of ever lighter blue tones in the left panel of
Fig.~\ref{fig:z02_Kip_Xabun}.  The temporal evolution of the abundance
stratification is more complex in diffusive models. The right panel of
Fig.~\ref{fig:z02_Kip_Xabun} illustrates that elemental diffusion is
effective enough that between all H-flashes, when deep convection
ceases, hydrogen diffuses to the top of the star. Hence, chemically
the H-flashing star appears alternately as a H-depleted or even a
helium star (during a flash) or a star with a thin essentially pure
hydrogen envelope. The thickness of the superficial hydrogen layer in
diffusive models, of the order $10^{-4} - 10^{-6}\,\mast$, is much
thinner than in the case if elemental diffusion is neglected.

\begin{figure}
\label{fig:z02_Surf_Xabun}
\centering
\includegraphics[width=0.445\linewidth]{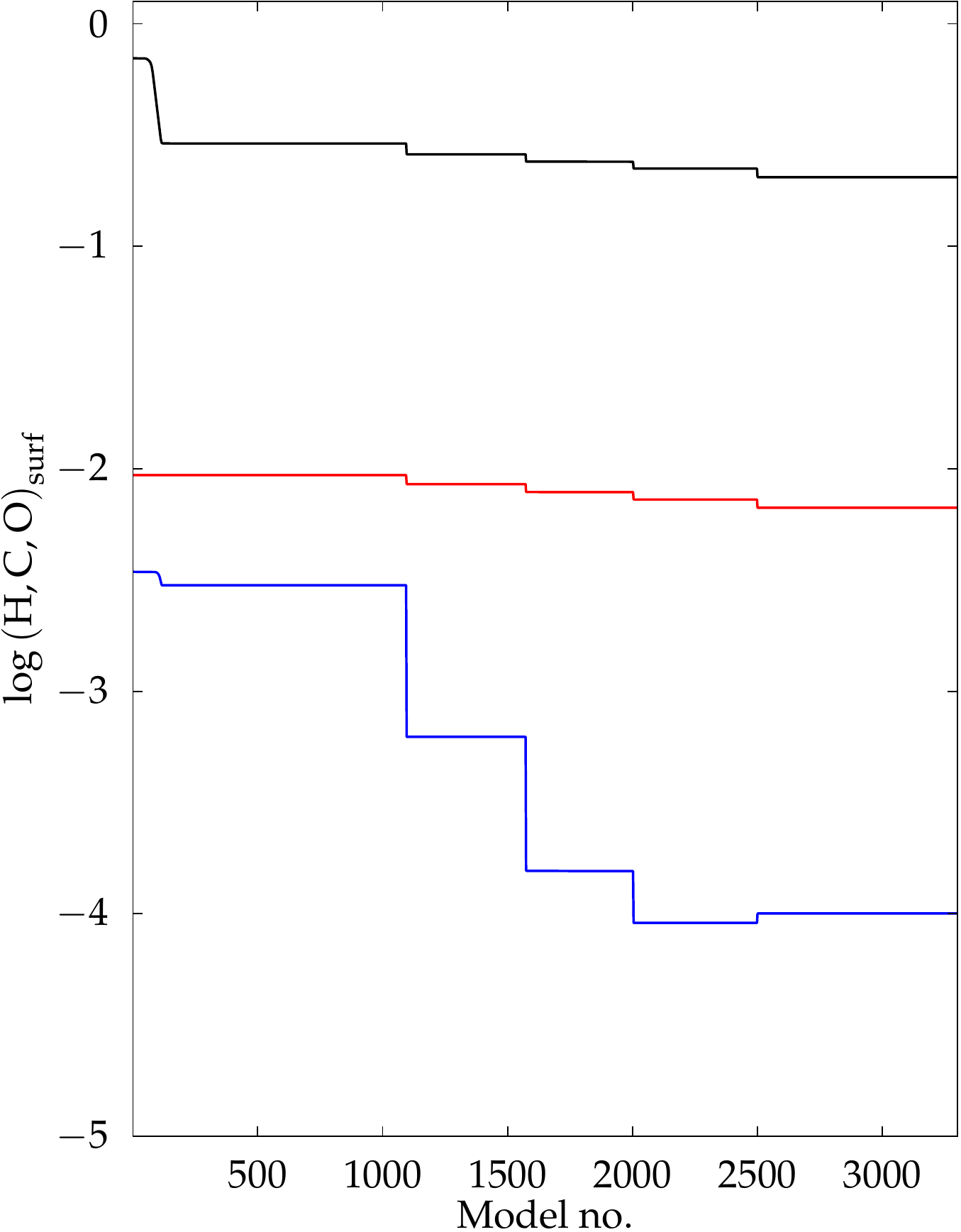}
\hspace{0.3cm}
\includegraphics[width=0.445\linewidth]{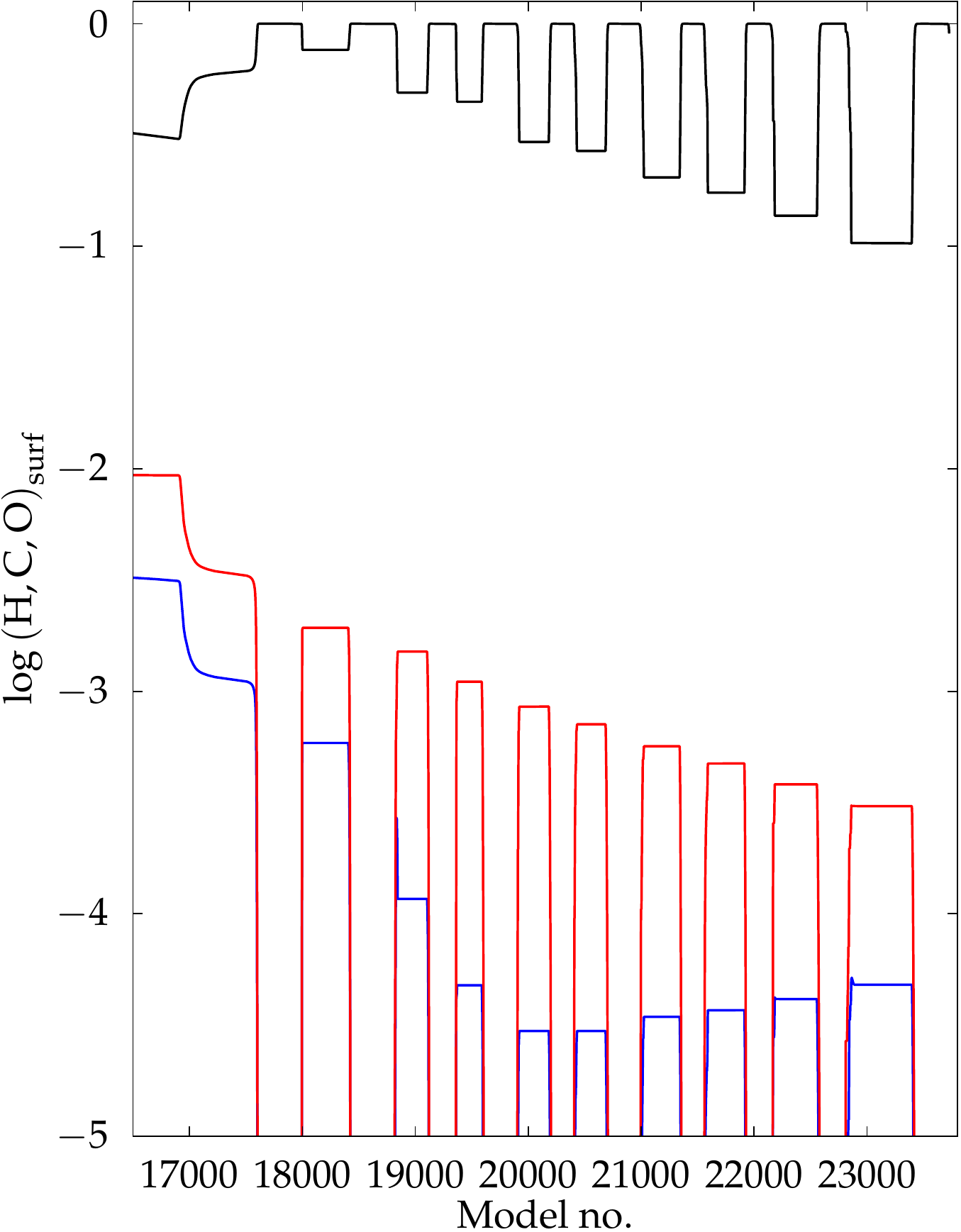}
\caption{Comparison of the temporal evolution of the surface
  abundances of hydrogen (black), $^{12}$C (blue), and $^{16}$O (red)
  of the $Z=0.02, 0.19\,\msol$ model series without diffusion (left
  panel) and including elemental diffusion (right panel).  }
\end{figure}

Evolution of the surface abundances of hydrogen (black), $^{12}$C
(blue), and $^{16}$O (red) are displayed for non-diffusive star models in
the left and for models including elemental diffusion in the right
panel of Fig.~\ref{fig:z02_Surf_Xabun}.

Evidently, the spectral appearance of old VLM stars, in particular
during the thermal-pulse episode, is different for diffusive and
diffusion-free models. In the latter case the hydrogen abundance
decreases monotonically in time; during each H-shell flash it drops
quickly~--~leading to a step in
Fig.~\ref{fig:z02_Surf_Xabun}. Hydrogen depletion starts already
during core hydrogen burning due to the star's fully convective
structure; once the star arrives on the white-dwarf cooling-track, it
evolves into an increasingly more pronounced hydrogen-deficient white
dwarf. Parallel to hydrogen, also the abundances of $^{12}$C and
$^{16}$O diminish on the surface, with the ratio C/O sinking from
about 0.36 to roughly 0.02 after the thermal pulses. Models with
diffusion present a less monotonous picture. During the H-shell
pulses, when the envelope is convective, the surface hydrogen
abundance is lower than in the diffusion-free models. As soon as the
envelope becomes radiative again~--~during most of the interpulse
phase~--~elemental diffusion strongly modifies the superficial
abundance structure. The envelopes get purified in the sense that the
superficial layers consist then essentially of pure hydrogen. Each
interpulse period is long enough for this purification to take
effect. Hence, either the thermally pulsing VLM stars are observable
as hydrogen-deficient, with their hydrogen-deficiency to grow with
advancing age and with decreasing luminosity (if the star is not on an
extended tour of the HR plane as a consequence of the H-shell flash),
or the VLM stars of a large range of luminosities, lying along the
white-dwarf cooling track, show a black-body spectrum with an overlaid
chiefly pure hydrogen line-spectrum. The temporal evolution of the
abundances of $^{12}$C and $^{16}$O show the mirror image of that of
hydrogen. During the phases of convective envelopes the C and O
abundances at the surface build up, during radiative phases of the
envelope these heavier elements diffuse inward and essentially
disappear from the surface. The general trend of reducing the C (at
least up to the $5^{\mathrm{th}}$ pulse) and the monotonous reduction
of O across all thermal pulses is functionally comparable to the
diffusion-free case. However, if elemental diffusion is accounted for,
the magnitude of reduction across the thermal-pulse phase is much
larger. Also the C/O ratio changes differently. The ratio starts at
about 0.35 before the thermal pulses set in; around the
$4^{\mathrm{th}}$ pulse C/O reaches a minimum at 0.04 to rise back to
about 0.16 when the star reaches its terminal cooling phase.

\section{The Physics of Hydrogen Shell Flashes}
Not long after the first encounter of H-shell flashes in VLM stars
\citep{Kippenhahn1968} the phenomenon was physically analyzed and
explained
\citet{Giannone1967}\sidenote[][-3.7cm]{ The violation of causality is
  only ostensible: \citet{Kippenhahn1967a} was the precursor paper to
  \citet{Kippenhahn1968}; the earlier paper computed the evolution of
  the initially $2.0\,\msol$ primary star of a close binary system to
  the point of onset of the secular instability of the
  hydrogen-burning shell when the primary was stripped down to
  $0.26\,\msol$. Weigert, working in the Kippenhahn group and having
  come across helium-shell flashes in intermediate mass stars earlier, was
  well prepared to apply his expertise to secular instabilities of
  \emph{hydrogen-burning} shells. }
expanding on the ansatz of \citet{Schwarzschild1965}.

\citet{Giannone1967} asked under what conditions a heat perturbation
continues to grow if it is applied to a thin nuclear burning
shell. The reaction of the nuclear active shell and the form of the
perturbations were highly abstracted, but the method allowed to
formulate simple analytical conditions that are to be met for
\emph{secular instability} to develop. Derived from the linear
stability analysis of the energy and the transport equation, the
following two criteria are necessarily to be fulfilled: \emph{The
  shell must be thin enough}:
\begin{equation}
  \label{eq:D1}
  D_{1}\equiv4\left[\alpha-\nabad\cdot\delta\right]\cdot\frac{\Delta r}{r}<1\,,
\end{equation}
 but at the same time, \emph{the shell must not be too thin}:
\begin{equation}
  \label{eq:D2}
  D_{2}\equiv\frac{8}{\Delta\ln T\cdot\varepsilon_{T}\cdot\left(\Delta
      L/L\right)}<1\,.
\end{equation}
\marginnote[-2cm]{ The pair $\{P,T\}$ is adopted as the
  thermodynamical basis; the characteristic exponents of the equation
  of state, $\alpha$ and $\delta$, are related to the parametrization
  $\rho\sim P^\alpha T^{-\delta}$. The variation of physical
  quantities $q$ across the nuclear burning shell are written as
  $\Delta q$. The rest of the symbols has canonical meaning, following
  closely the usage in \citep{kw}.  } 

The condition of the shell to be thin
enough~--~Eq.~(\ref{eq:D1})~--~ensures that the pressure reaction
during expansion (after some heat perturbation be applied to the
nuclear burning shell) remains sufficiently weak so that the
temperature in the shell to continues to grow. Equation~(\ref{eq:D2})
on the other hand must ensure that the shell is not too thin in order
for the heat perturbation to remain contained in the shell. This can
be accomplished either by a sufficiently strong nuclear burning
(manifested through the magnitude of $\varepsilon_T$ and $\Delta L$,
the luminosity contrast across the shell) or then by sufficiently weak
energy leakage, measured by the temperature contrast $\Delta \ln T$
through the thin nuclear-burning shell.

Both conditions, $D_1$ and $D_2$, assume homologous motion throughout
the nuclear-burning shell. Numerical models show that actually the
hydrogen burning shell is \emph{the} region of the star where
homologous change is least satisfied ($0.7 \lesssim -4*\delta\ln r /
\delta\ln P\lesssim 0.9$)
\sidenote[][]
{$\delta\ln Q \equiv \ln Q(m,t_1) - \ln Q(m,t_0)$, \\ 
  \parindent=0pt
  i.e. a Lagrangian temporal difference of quantity $Q$. 
},
nonetheless the deviations from homology are small enough 
for the instability conditions to remain predictive.

\begin{figure}
	\label{fig:hpulse_discriminants}
	\includegraphics{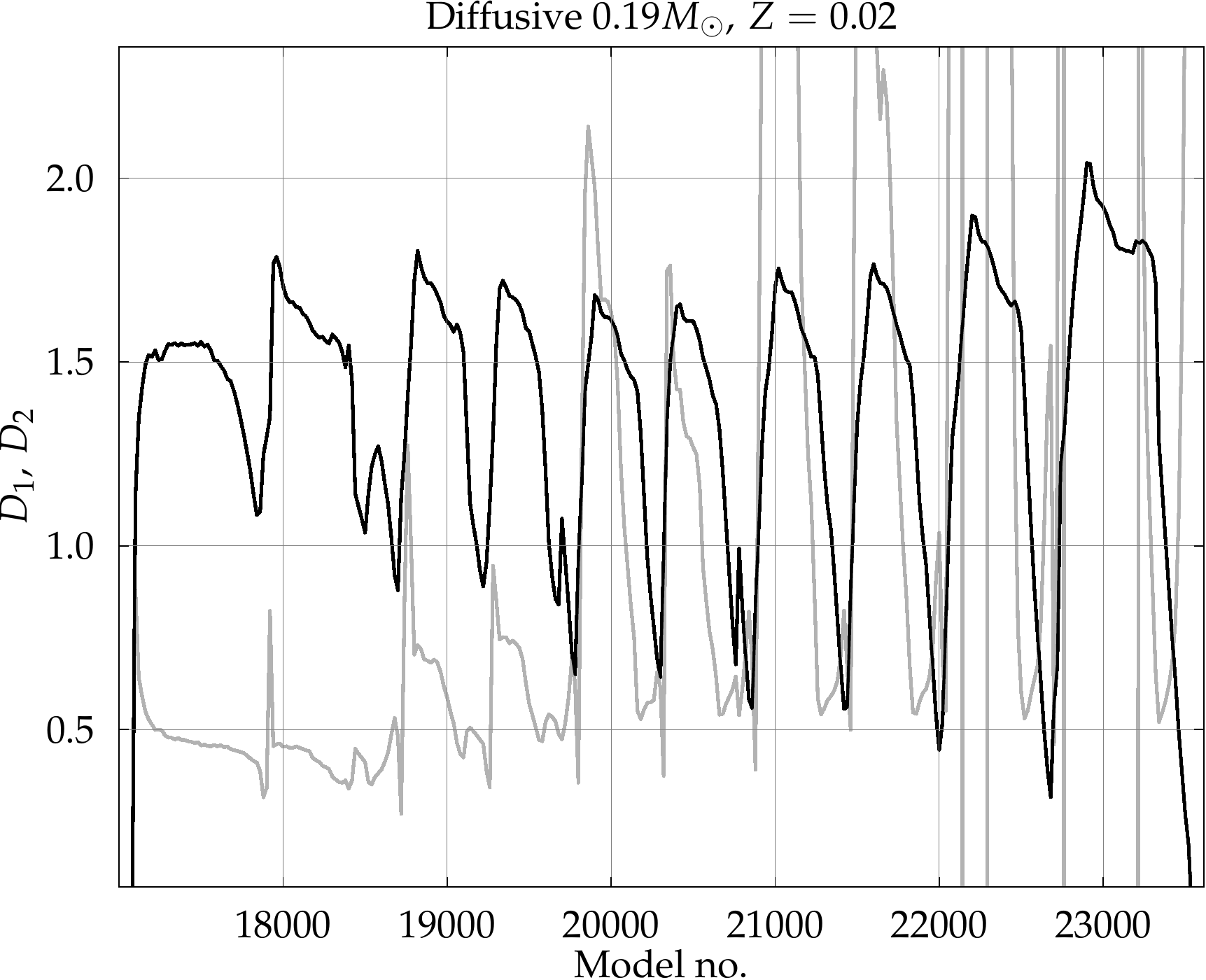}
	\caption{Temporal evolution of the stability discriminants
          $D_1$ (black line) and $D_2$ (gray line) during the H-shell
          flash phase of the diffusive $0.19\,\msol,\,Z=0.02$ model
          sequence. The size of the nuclear burning shell was
          determined at $\varepsilon=5$~erg/g/s level.  }
\end{figure}

The variation of the determinants $D_1$ (black) and $D_2$ (gray) as
determined in diffusive $0.19\,\msol,\,Z=0.02$ models during their
thermally pulsing phase are presented in
Fig.~\ref{fig:hpulse_discriminants}. The behavior of the two
quantities is representative for all other cases we studied for this
exposition. The actual magnitude of the discriminants, in particular
of $D_2$, depends on the details of how the hydrogen-burning shell is
defined. For Fig.~\ref{fig:hpulse_discriminants}, the region with
$\varepsilon_{\mathrm{nuc}} > \varepsilon_{\mathrm{crit}} \equiv
5$~erg/g/s is assumed to measure the extension of the hydrogen-burning
shell. The physical quantities entering the stability criteria are
evaluated at the bottom of the shell as this is the dominant location
for the analysis. The thermodynamic quantities in $D_1$ and
$D_2$ do not vary strongly across the shell, this applies in
particular for the quantities entering $D_1$; even the measure $\Delta
r / r$ is robust since the nuclear-burning shell is geometrically
thin. On the other hand, all the quantities in the denominator of
$D_2$ in Eq.~(\ref{eq:D2}) depend on varying degree on the particular
choice of $\varepsilon_{\mathrm{crit}}$.

The variation of $D_1$ during the flash cycle is dominated by the
quantity $\Delta r / r$ of the nuclear burning shell, which can also
be discerned clearly in the Kippenhahn diagrams of
Fig.~\ref{fig:z02_Kip_NucBurn}. It is also the monotonously shrinking
value of $\Delta r / r$ which lets $D_1$ eventually to tend to zero.
The variability of $1/\Gamma_1\equiv\alpha-\nabad\cdot\delta$ is small
and is not important for the functional behavior or the amplitude of
the variability of $D_1$. In particular, $\Gamma_1$ of the
hydrogen-burning shell does not change systemically during the
evolution through the thermal-pulse phase.

The magnitude, but not the amplitude of variability of $\varepsilon_T$
is important for instability in condition $D_2$. The range $12
\lesssim\varepsilon_T\lesssim 17$ as measured in all models of all
analyzed sequences points unambiguously at \emph{CNO burning as the
  dominating nuclear energy source} in the shell. In particular,
$\varepsilon_T$ stays that high at the end of the thermal pulsing
phase also in the least massive studied models which we computed. Put
otherwise, the young white dwarfs turn secularly stable before their
ever weakening nuclear burning becomes $pp$-burning dominated.  It is
the cyclic variation of $\Delta L/L$, evaluated at the edges of the
nuclear burning shell, which dominates variability of $D_2$.
Additionally, the long-term change of the temperature contrast,
$\Delta \ln T$, across the shell contributes to the ever rising trend
of $D_2$ which underlies its cyclic variability.

For the model sequence under consideration in
Fig.~\ref{fig:hpulse_discriminants}, criterion $D_1$ in particular is
observed to not develop a sufficiently deep minimum at the first
thermal pulse to correctly predict instability (as seen in the full
evolution computations) but it does well for the other eight pulses
which were numerically tracked in the full nonlinear evolution
computations. Changing the threshold value of
$\varepsilon_{\mathrm{crit}}$ can remedy this problem, but at the same
time this renders the variations of $D_2$ wilder. In any case, form
and amplitude of the \emph{variation} of $D_1$ and $D_2$ during the
episode of the thermal pulses remain robust and therefore the criteria
are \emph{helpful to understand} the physical processes at work during
the secular instability. The sensitivity of the numerical values of
$D_1$ and $D_2$ on their implementation renders them unpractical to be
used directly during evolutionary computations, e.g. for automatic
time-stepping choices in stellar-evolution codes.
 
All in all, we find that for sufficiently massive VLM stars, usually
with $\mast \gtrsim 0.32 \msol$, have sufficiently fat hydrogen-burning
shells during their post red-giant evolution so that the nuclear
burning region never becomes secularly unstable. For the mass range
roughly $0.2 - 0.35 \msol$, the hydrogen-burning shell can get thin
enough to trigger one or more hydrogen flashes through the cyclic
shrinking of discriminant $D_1$.  The secular instability dies out
once the shell weakens enough, a process sped up by the enhanced
outward-eating of the shell during the thermal pulses, so that it
finally cannot retain enough of the heat perturbation in the shell to
feed the thermal runaway; i.e. it is discriminant $D_2$ which
quenches the instability. The low-mass boundary for secular
instability is again defined by discriminant $D_2$ which measures the
stabilization via a too weak luminosity and temperature contrast
across the hydrogen-burning shell.

\begin{margintable}
\begin{tabular}{l c l}
\toprule
$(M_{\mathrm{lb}} - M_{\mathrm{ub}})/\msol$ &  Diffusion & $Z$ \\
\midrule
0.19  - 0.32  & $\times$     & 0.02  \\
0.25  - 0.35  & $\times$     & 0.001 \\
0.175 - 0.36  & $\checkmark$ & 0.02  \\
\bottomrule
\end{tabular} 
\end{margintable} 
The table on the margin lists the lower ($M_{\mathrm{lb}}$) and upper
mass ($M_{\mathrm{ub}}$) boundaries in between of which hydrogen
flashes were encountered in our MESA modeling. The column labeled with
``Diffusion'' indicates if elemental diffusion neglected ($\times$) or
was accounted for ($\checkmark$). The last column lists the
heavy-element abundance of the model sequence. The dependency of the
mass range of the secular instability on metallicity appears to be
stronger than on including/neglecting elemental diffusion in stellar
modeling. This observation, before being adopted as a general rule
needs support by a much broader computational survey of the relevant
parameter space.

\section{Wrap-up}
\newthought{Secular instability of hydrogen-burning shells} can be run
into by VLM stars under suitable conditions; this has been known for a
long time, starting with the findings of \citet{Kippenhahn1968}. In
the present exposition, the phenomenon was revisited resorting to
detailed but exemplary single-star models computed with the MESA
stellar evolution code. The results were used to collect pertinent
data to foster our understanding of the secular instability of the
hydrogen-burning shell.

Above all, hydrogen-shell flashes are a robust phenomenon, they occur
for a broad range of chemical compositions, with and without elemental
diffusion; hence the inflicted reactions of the stars should be
observable at appropriate epochs of the cosmic evolution.  Depending
on the particular microphysics, i.e. chemical composition, mixing
processes and the like, the phenomenology of the thermal flashes and
the stellar mass range that develops secular unstable hydrogen-burning
shells varies. Very roughly, VLM star models with secularly unstable
hydrogen shells can be found in the mass range roughly between $0.2$
and $0.35\msol$.

\newthought{Astronomy in the aging Galaxy} is potentially less boring
than advocated in the past \citep[e.g.][]{Adams1997}.  With regard to
the terrestrial civilization, however, \emph{single-star evolution}
with its thermally pulsing VLM objects is not relevant. It takes the
pertinent VLM stars about $ 1- 10 \times10^{11}$~years to develop
their secularly unstable H-burning shells.  By then, the sun will have
evolved into a faint, cool white dwarf \citep[e.g.][]{Sackmann1993}
probably with only a fragmentary planetary system left, and in
particular without a habitable earth. Hence, human civilization on
earth is going to miss the light-show that should develop around the
Galaxy's retirement from its stelliferous era.

Indifferent to the human possibilities to witness single VLM-star
H-flashes, we find that when using the initial mass function (IMF) of
\citet{Chabrier2003} for Z=0.02 mass bounds for secular instability,
that about 33\% of all stars with $\mast < M_{\mathrm{ub}}$
will pass through H-shell flashes when the stars with $\mast =
M_{\mathrm{ub}}$ start flashing. For the case of $Z=0.001$, the
fraction of H-flashing stars drops to about $19\%$. These numbers are
lower boundaries since the star-formation history was no single epoch
event. Down to at least about $0.01 \msol$ the IMF has a negative
slope so that the number of H-flashing candidates is huge in the
Galaxy. Therefore, the Milky Way should appear quite variable for some
time in the Cosmic future, even when it will be made up exclusively by
degenerate left-over stellar bodies. 

Hence, once the flashing young-white dwarfs brighten up the Milky Way,
at least the more metal-rich members are prone to become born-again
red giants. A future civilization, even without a heavy
stellar-physics history backpack will have another chance to enjoy
red-giant like objects; this without ever having seen the by then
extinct kind that comes about in our traditional way: As stars
evolving away from the main sequence on their way to try to burn
helium.

In a population of VLM stars with some spread in stellar formation and
a mixture of masses, a future observer of the Galaxy should encounter
a mixture of hydrogen-deficient (typically brighter, because the
elemental mixing happens during the flash cycle when the VLM stars
brighten considerably) and helium-white dwarfs with shallow but very
pure hydrogen layers (upon their crossing the instability regions of
white dwarfs, asteroseismology~--~should it still be pursued
then~--~can be predicted to yield measures of superficial H-layer
thicknesses of the order of $10^{-3}-10^{-5}\,\mast$).

The chances for our civilization to witness thermally pulsing single
VLM stars are note completely bleak though: The Galaxy might even
currently harbor a few candidates.  \citep{Kilic2007} reported of
numerous low-mass white dwarfs for which no indications of a close
companion could be observed; the authors argued that
super-metal-rich single stars might suffer from enhanced mass-loss
when evolving up the \first~giant branch that some stars of
the lower-mass fraction do not reach He-ignition and leave the giant
branch prematurely to cool then as low-mass helium white dwarfs in less
than a Hubble time.
\marginnote[-2cm]{There is observational evidence from the populations
  of red giants in stellar aggregates such as globular clusters and
  the local neighborhood in the Galaxy that red-clump giants are
  underabundant or even missing as [Fe/H] increases. See
  \citet{Kilic2007} for references.}
However, it seems that only about 50\% of $0.4\msol$ white dwarfs
\citep{Kilic2007}, a number changing to 75\% according to the sample
analyzed in \citep{Brown2011}, are thought to emerge from the
single-star channel. Towards even lower masses, the fraction of pure,
single stars sinks rapidly and extremely-low-mass (ELM) white dwarfs
with $\mast < 0.2\msol$ are expected to be exclusively produced by
interactions in close-binary systems as it is illustrated in the next
section. Since only white dwarfs with masses approximately $0.2 - 0.35
\msol$ develop secularly unstable H-burning shells, and those close to
the low-mass boundary are the most ``violent'' and hence the most
flamboyant ones, only a tiny fraction, if at all, of low-mass helium
white dwarfs remain as candidates at the end of the exclusion process.

\newthought{In close binary stars} in which mass is abstracted from
the primary component via evolution scenario A or B, the donor can
turn into a low-mass star unable to ignite core helium burning.
Within a fraction of a Hubble time, such stars can end up as low-mass
or even ELM helium white dwarfs. The initial \emph{modeling} with
evidence for H-shell flashes was found just in this framework
\citep{Kippenhahn1968}.

Hints at the binary channel to produce low-mass helium white dwarfs
come already from the entrance, i.e. from VLM \emph{subdwarfs} and
\emph{proto white-dwarfs} in close binary systems, which were
discovered and monitored in the recent past. \citet{Maxted2012}
reported on the mass determinations of the components of the
double-lined eclipsing system J0247-25 and their Fig.~5 presents the
distribution of nine yet calibrated low-mass binary components on the
HR plane and their relation to stellar evolution tracks adopted from
the earlier modeling literature. The photometric space mission
\emph{Kepler} yielded recently a very intriguing low-mass star that
must have shortcut the ascent of the \first~giant branch and which is
presently observed to cross the classical instability strip and is
observable as an RR Lyrae~--~type pulsating
variable\sidenote[][-2cm]{The system RRLYR-02792 is a detached, double-line
  eclipsing binary with the primary component, the pulsating star,
  being observed with a mass of $0.26 \msol$. The secondary component
  is attributed a mass of $1.67 \msol$. The two stars orbit each other
  every $15.24$~days; the pulsation period of the primary star is
  $0.627$~days. About ten years of monitoring shows the period to
  clearly decrease ($\dot{P}=-8.4\times10^{-6}$~days/year) indicating
  a blueward evolution of the star~--~just as expected by the invoked
  scenario.}
\citep{Pietrzynski2012}.
The low-mass pulsator should evolve into a hot white dwarf within
about the next $10^7$~years and, based on the estimated mass, pass
through at least one hydrogen-shell flash.

The large-scale sky survey SDSS reveals the ubiquity of ELM white
dwarfs \citep[e.g.][]{Brown2012,Kilic2012} in the Galaxy. Even the
first three examples of old ELM
white dwarfs\sidenote[][+0.5cm]{SDSS J184037.78+642312.3 \\
\parindent=5pt  
SDSS J151826.68+065813.2 \\ 
\parindent=5pt
SDSS J111215.82+111745.0 \\
}
which have cooled down into the ZZ Ceti instability strip were
recently reported by \citet{Hermes2012,Hermes2013}.  Depending on the
richness of the eventually recoverable frequency spectra of the
pulsators an asteroseismic probing of the thickness of their
hydrogen-rich surface layer can be anticipated.

\bigskip

Observational evidence of ELM white dwarfs in double
degenerates~--~either in double white-dwarf systems (DWD) or in
white-dwarf~/~neutron-star pairs, in particular if the neutron star is
a millisecond pulsar (MSP), has amassed over the past few years
\citep[e.g.][and references therein to earlier
achievements]{Kilic2012,Marsh2011}.
 
\newthought{DWD systems} are at the center of considerable research
activity since, if they are sufficiently tightly bound and can merge
within a fraction of the age of the Cosmos, time and again they are
traded as candidates for type Ia supernovae (if the components are CO
white dwarfs, i.e. the components are massive enough) and sources of
gravitational waves, which might soon become detectable by the
detectors of the LIGO/VIRGO experiments.

The binary nature of low-mass white dwarfs has been observationally
established in the outgoing 20th century, starting with
\citet{Marsh1995} (finding white-dwarf masses of $\sim0.3-0.4\, \msol$) and was
thereafter supported and extended (also towards lower white dwarf
masses ; i.e. $\lesssim0.2\, \msol$ , see references in
\citet{Kaplan2012} by larger samples collected in surveys such as the
SDSS \citep[e.g.][]{Kilic2011}. DWD systems having endured multiple
common-envelope phases might explain the very short-period systems of
the extreme kind such as
HM Cnc\sidenote{also known as RX J0806.3+1527, an X-ray source
  discovered with ROSAT in 1999. The binary system consists of two
  white dwarfs, each of about $0.5\,\msol$ with an orbital
  period of 5.4~minutes!
} 
with mass transfer during the inspiraling of the two white dwarf
components. Eventually, such binary systems might evolve into helium
mass-transferring cataclysmic AM CVn stars \citep{Kaplan2012}.

ELM white dwarfs are predicted to possess stably burning H envelopes
($M_{\mathrm{env}}\sim10^{-3}-10^{-2}\msol$ ) which let them stay
bright for $\mathcal{O}$(Gyr) \citep{Serenelli2002, Panei2007}; this
can be understood from the fact that no hydrogen flashes can occur
which quickly reduce the H-rich envelope mass. Apparently, for the DWD
system NLTT 11748, the geometric measurement of the radius of an ELM
candidate was possible \citet{Kawka2009}, yielding $0.04\, \rsol$ for
an $0.15\, \msol$ He white dwarf; a value which appears to be
compatible with a thick stably burning hydrogen envelope.

With respect to cooling ages and possibly other observables, we expect
a split-up of the population of low-mass white dwarfs: The mass range
of thermally pulsing low-mass helium white dwarfs is expected to have
thin H-rich/H-pure envelopes and short cooling ages; on the other hand
the thermally stable VLM/ELM He white dwarfs that burn hydrogen stably
can live with thick H-rich envelopes and sustain significant
luminosity over long times and hence appear older at a given
luminosity.

\newthought{MSP systems} constitute one group of double degenerate
binary systems that contain a VLM white dwarf with $\mast \lesssim
0.4\,\msol$ and a neutron star with mass of the order of
$1.5\,\msol$. Due to its very low mass, the white dwarf is thought to
have a helium core with a thin hydrogen blanket.  The neutron star is
thought to have been spun up through angular-momentum transfer during
earlier mass-exchange episodes. Therefore, in this framework the MSP
birth epoch is taken to coincide with the start of the "cooling" of
the low-mass white dwarf that emerged from the mass-loss
episode. Hence, the cooling time (the time from starting on the
cooling sequence of He white dwarfs) and the spin-down age of the MSP
must coincide. A problem that hampered MSP understanding in the past
was the incompatibility of the model-derived age of the white dwarf
component and of that of the MSP.

As discussed in \citet[][and references therein]{Althaus2001}, and
reiterated by \citet{Benvenuto2005}, who also computed the orbital
evolution of the binary system, the inclusion of elemental diffusion
in modeling the white dwarf evolution proved essential to reduce the
cooling age of the white-dwarf companion of MSPs and to reconcile
their ages. If white dwarfs harbor thick envelopes, such as they
develop in diffusion-free models, the degenerate stars cool too slowly
since hydrogen burning can maintain the star's luminosity for a too long
period of time. On the other hand, accounting for elemental
diffusion allows already the pre-white dwarfs to burn their hydrogen
more efficiently due to the purified burning shells.
Since hydrogen flashes are encountered also in white-dwarf models in
which elemental diffusion was accounted for, this process opens
additional ways to effectively reduce the amount of superficial H-rich
material: Hydrogen flashes rapidly burn H to He and reduce the
thickness of the H-rich envelope from the bottom. Furthermore, if the
flash forces the star to puff up and to expand possibly beyond its
Roche radius the then starting rapid mass-loss can reduce the
thickness of the H-rich envelope by stripping it from the top. Any of
these processes forces a H-flashing star to further speed up its
cooling because it has less fuel to support its nuclear furnace
\citep[e.g.][]{Althaus2001}. \citet{Ergma2001} contemplated the
possibility of irradiation by the neutron star to play a potentially
significant r{\^o}le to chip away hydrogen-rich superficial matter
from the white dwarf when it is inflated during a hydrogen flash.

Since not all VLM/ELM helium white dwarfs undergo H-flashes, they only
contribute partially to the solution of the clocking problem in MSPs
which was discussed above.  Hence it should be very interesting to
strive for large enough samples of MSPs with calibrated white-dwarf
companion masses. In particular around the boundaries that separate
the H-flashing domain from the high- and low-mass stably burning
VLM/ELM white dwarfs, an enhanced age spread should be observable. The
data might be useful to scrutinize the involved microphysics that
determines structure and stability of the H-burning shell. If VLM/ELM
white dwarfs in MSP systems even happen to fall into one of the white
dwarfs' instability strips, asteroseismology will be a ready tool to
probe the envelope structures and might supply an independent measure
of the hydrogen-layer thickness.

\newthought{An outlook on more massive stars}, still of the low-mass
kind but having entered helium burning at some stage during their
evolution, reveals that they too can run into unstable
hydrogen-burning shells under suitable circumstances during the
post-AGB stage. One mechanism was promoted by \citet{iben1986a}: Young
white dwarfs that left the AGB during either He-shell or early
H-shell burning manage to contaminate their essentially pure helium
mantle with $\H1$ and $\C12$ through chemical diffusion. If the
accumulation of C, N, and H become favorable in the right temperature
window, unstable CNO-burning can emerge. The lightcurve produced by
the resulting thermal flash appeared to be reminiscent of very slow
novae so that the authors referred to the phenomenon as a
\emph{self-induced nova}. \citet{MillerBertolami2011} who computed
more self-contained white-dwarf models with proper antecedents
revisited the scenario, rebaptizing it to the more designative
\emph{diffusion-induced nova}, to scrutinize it with a wider range of
constitutional parameters. The stability behavior of the CNO-burning
shell was paid attention to by applying the formalism which was used
earlier on by \citet{Yoon2004} who relied on the local
linear-stability ansatz of \citet{Giannone1967}.  The secular shell
instability that leads to diffusion-induced novae can be understood
along the same line of arguments as presented in this exposition.

\vskip 1cm

\newthought{Acknowledgments} This exposition relied heavily on NASA's
Astrophysics Data System Bibliographic Services. The value of the
efforts of Bill Paxton and the MESA community to make available to the
student of the stars, irrespective of affiliation and status, an
open-source one-of-a-kind stellar evolution code such as MESA can
hardly be overstated. Hideyuki Saio and Leandro Althaus generously
donated of their time to read through the exposition with critical
eyes.

\vskip 1.5cm
\section{Appendix: Star Modeling}
The stellar models referred to and analyzed in this exposition were all
computed with the MESA code in its version 4298
\citep[cf.][for a general description of the MESA project]{Paxton2011,
Paxton2013}. 
\marginnote{For the opacity data, \texttt{gn93} and
  \texttt{lowT\_fa05\_gs98} tables at low temperatures were requested
  in MESA. The EoS was computed with \texttt{macdonald} data for its
  smoothness.}
Convection was treated with the Henyey scheme assuming an ad hoc
mixing-length of 1.8 pressure scale-heights. The choice $Z=0.02,
X=0.7$, as an example of PopI abundances, allowed the sequence of
homogeneous models to start from initial models located close to the
ZAMS. For the exemplary PopII abundances, chosen as $Z=0.001,
X=0.757$, the evolution computations were started with homogeneous
pre-main-sequence polytropes, which first contracted along the Hayashi
track onto the ZAMS to be then followed up to their terminal cooling
stage as helium white dwarfs.

Elemental diffusion in MESA is treated for the species under
consideration by solving Burgers' equation in parallel to the stellar
structure/evolution equations, the implementation in MESA is
comparable to that described in \citet{Althaus2000}.
\marginnote[-1.5cm]{\texttt{do\_element\_diffusion = .true.} was set in
  the \texttt{\&controls} section of the inlist; we adopted the
  default settings from \texttt{star\_defaults.dek} with 5 species
  classes that were used for the diffusion computations.    
         }

\vskip 2.0cm

\bibliographystyle{aa} 
\bibliography{StarBase}

\end{document}